# Magnetic Switching in Monolayer 2D Diluted Magnetic Semiconductors via Spin-to-Spin Conversion

*Siwei Chen, Zitao Tang, Mengqi Fang, Rui Sun, Xiaotong Zhang, Licheng Xiao, Seyed Sepehr Mohajerani, Na Liu, Yuze Zhang, Abdus Salam Sarkar, Dali Sun, Stefan Strauf, and Eui-Hyeok Yang\**

S. Chen, Z. Tang, M. Fang, A. S. Sarkar

Department of Mechanical Engineering, Stevens Institute of Technology, Hoboken, NJ 07030, USA

R. Sun, X. Zhang, D. Sun

Department of Physics and Organic and Carbon Electronics Laboratories (ORaCEL), North Carolina State University, Raleigh, NC, 27695 USA

L. Xiao, S. S. Mohajerani, N. Liu

Department of Physics, Stevens Institute of Technology, Hoboken, NJ 07030, USA

Y. Zhang

Department of Chemical Engineering and Materials Science, Stevens Institute of Technology, Hoboken, NJ 07030, USA

S. Strauf

Center for Quantum Science and Engineering

Department of Physics, Stevens Institute of Technology, Hoboken, NJ 07030, USA

E. Yang

Center for Quantum Science and Engineering

Department of Mechanical Engineering, Stevens Institute of Technology, Hoboken, NJ 07030, USA

eyang@stevens.edu

**Abstract**

The integration of two-dimensional (2D) van der Waals (vdW) magnets with topological insulators or heavy metals holds great potential for realizing next-generation spintronic memory devices. However, achieving high-efficiency SOT switching of monolayer vdW magnets at room temperature poses a significant challenge, particularly without an external magnetic field. Here, we show field-free, deterministic, and nonvolatile SOT switching of perpendicular magnetization in the monolayer, diluted magnetic semiconductor (DMS), Fe-doped $MoS_2$ (Fe:$MoS_2$) at up to 380 K with a current density of ~7 × $10^4$ A $cm^{-2}$. The in situ doping of Fe into monolayer $MoS_2$ via chemical vapor deposition and the geometry-induced strain in the crystal break the rotational switching symmetry in Fe:$MoS_2$, promoting field-free SOT switching by generating out-of-plane spins via spin-to-spin conversion. An apparent anomalous Hall effect (AHE) loop shift at a zero in-plane magnetic field verifies the existence of z spins in Fe:$MoS_2$, inducing an antidamping-like torque that facilitates field-free SOT switching. A strong topological Hall effect (THE) was also observed, attributed to the interfacial Dzyaloshinskii-Moriya interaction (DMI), reducing the energy barrier for SOT switching. This field-free SOT application using a 2D ferromagnetic monolayer provides a new pathway for developing highly power-efficient spintronic memory devices.

## 1. Introduction

Two-dimensional (2D) van der Waals (vdW) magnets have emerged as promising contenders in spintronics, primarily owing to their heightened spin transparency and spin injection efficiency, which results from the absence of dangling bonds and decreased atom intermixing near the interface at the heterostructure surface.[1–5] The spintronics of 2D vdW magnets are particularly beneficial because of their strong perpendicular magnetic anisotropy (PMA),[6–11] paving the way toward versatile manipulation of magnetic ordering via current-induced spin-orbit torque (SOT). SOT switching involves applying a current pulse that induces torque on the magnetization of materials through spin–orbit interactions, efficiently altering the magnetic states of magnetic layers.[12] The spin-orbit interaction is facilitated by integrating 2D magnets with materials exhibiting robust spin-orbit coupling (SOC), such as topological insulators (*e.g.*, $WTe_2$, $Bi_2Te_3$) or heavy metals (*e.g.*, W, Ta, Pt).[8,13–15] The synergy between these materials and vdW magnets may significantly enhance the SOT efficiency, attributed to the inherently low magnetic damping in vdW magnets.[8,16,17]

To achieve high efficiency in SOT switching, it is crucial to select high spin transparency materials and minimize the thickness of the magnetic layer to reduce magnetic damping and saturation magnetization.[18–20] Recently, the use of thin $BaPb_{1-x}Bi_xO_3$ and $Fe_3GaTe_2$ heterostructures showed 'field-assisted' SOT switching with current densities on the order of 3 × $10^6$ A $cm^{-2}$ at 300 K,[13,21] with the application of in-plane magnetic fields. However, the ability to achieve SOT switching without an external magnetic field is essential for practical applications. To achieve field-free SOT switching, approaches such as lateral asymmetry,[22] in-plane exchange bias fields,[23] interlayer exchange coupling[24] and interfacial Dzyaloshinskii-Moriya interaction (DMI)-induced symmetry breaking[25] have been explored. A major advancement was reported using the low-crystal-symmetric material $WTe_2$ combined with room temperature vdW magnets, where an unconventional out-of-plane anti-damping torque facilitates field-free switching.[11,26] Field-free SOT switching with $WTe_2/PtTe_2/CoFeB$ and $WTe_2/Fe_3GaTe_2$ devices was achieved at a current density of $2.25 \times 10^6$ A cm$^{-2}$ at 300 K [11] and $2.23 \times 10^6$ A cm$^{-2}$ at 320 K,[10] respectively, while the Curie temperature of $Fe_3GaTe_2$ was 330 K. These vdW magnets (*e.g.*, $CrTe_2$, $Fe_3GeTe_2$ and $Fe_3GaTe_2$) are synthesized via flux crystal growth [7,27,28] or molecular beam epitaxy,[19] often followed by mechanical exfoliation in a glove box.

Furthermore, thinning these 2D magnets may enhance the system's energy efficiency, as entire magnetic layers interact closely with the spin-accumulated interface[17,29] and contribute to lowering the energy barrier for magnetization switching by reducing the saturation magnetization.[9,19] However, the Curie temperatures of $CrTe_2$ and $Fe_3GaTe_2$ drop to 224 K and 200 K, respectively, when reduced to monolayer thicknesses.[27,30] Furthermore, the surface of these materials is quickly oxidized when exposed to air,[19,31,32] which alters the material's electronic structure and introduces defects and structural irregularities, impairing its performance.[19,27,31,32] This thickness limitation presents a significant challenge for low-current-density, field-free SOT switching. Conversely, substitutional doping of monolayer transition metal dichalcogenides (TMDs), including $MoS_2$ and $WSe_2$, synthesized via chemical vapor deposition (CVD), thus providing ultimate scalability,[33–36] enables the formation of 2D diluted magnetic semiconductors (DMS).[37–39] These DMSs exhibit ferromagnetic properties above room temperature, even at monolayer thicknesses,[37–39] which makes them ideal candidates for SOT materials at the true 2D limit (i.e.*,* atomically thin monolayers).

Here, we show an ultralow current-density, field-free, deterministic SOT switching of perpendicular magnetization in a CVD-grown, monolayer Fe:$MoS_2$ transferred onto a Pt Hall bar. We demonstrate the PMA in Fe:$MoS_2$ by measuring the anomalous Hall effect (AHE) in Fe:$MoS_2$/Pt heterostructure up to 380 K. We also show that field-free, perpendicular magnetization switching by SOT occurs only when the current flows along the zigzag direction of Fe:$MoS_2$ due to strain-induced crystal asymmetry. We demonstrate that the rotational symmetry breaking in Fe:$MoS_2$, promoting field-free SOT switching by generating out-of-plane spins via spin-to-spin conversion. To further explore the origins of field-free and low-current-density SOT switching, we characterize the current-induced AHE loop shift and the temperature-dependent topological Hall effect (THE).

## 2. Results and Discussion
### 2.1 Materials and device characterization

Monolayer Fe:MoS$_2$ was grown via chemical vapor deposition (CVD) (Figure S1). The detailed synthesis and magnetic characterization processes, which showed that the material was a room-temperature DMS, are described elsewhere.[37,39] A schematic of the crystal structure of monolayer Fe:MoS$_2$ is shown in Figure 1A (left), illustrating that Fe atoms substitute for Mo sites within the lattice. The field-free SOT switching of perpendicular magnetization can be facilitated by utilizing low-symmetry crystals with, at most, one mirror symmetry.[7,10,11,40,41] While pristine MoS$_2$ exhibits threefold rotational symmetry, the uniaxial strain applied along the y- or x-axis disrupts the rotational symmetry in the hexagonal lattice,[42,43] retaining only the y-z mirror plane symmetry, σ$_v$ (yz) (z represents the out-of-plane axis), as shown in Figure 1A. The step height of the Pt film, as illustrated in Figure S2A, induces uniaxial strain ($\varepsilon$) in the monolayer Fe:MoS$_2$ perpendicular to the edge of the Pt Hall bar.[44–47] Optical images of a typical Fe:MoS$_2$/Pt heterostructure are shown in Figure S2B,C.

To determine the lattice orientation of Fe:MoS$_2$, thus controlling the strain direction, we utilized the edge morphology of triangular island single crystals,[48] as shown in Figures 1B and S3, with layers of Pt and FeMoS$_2$ marked. Figure 1A (right) shows a schematic representation of the strained Fe:MoS$_2$ lattice structure along the armchair and zigzag directions, effectively modulating the crystal symmetry. Previous studies have shown that, in monolayer MoS$_2$, covalent bonding between Mo and S is very sensitive to in-plane uniaxial strain due to the strain-induced change in the coupling between the Mo atom $d$ orbital and S atom $p$ orbital, which can be probed using angle-resolved polarized Raman spectroscopy. For MoS$_2$, in the absence of strain, the $E_{2g}^1$ mode intensities show negligible change.[42,43] This trend persists in monolayer Fe:MoS$_2$ transferred onto a flat SiO$_2$ surface, for which the $E_{2g}^1$ mode intensity difference is less than 5% (Figure S4B). To confirm the alteration in crystal symmetry, we analyzed the angle-resolved polarized Raman spectra of monolayer Fe:MoS$_2$ transferred onto a Pt Hall bar. The polar Raman laser spot is indicated with the white arrow in Figure 1B. Figures 1C, S4a and S4b present their contour maps, where two dominant peaks at 384 cm$^{-1}$ and 403

cm$^{-1}$, corresponding to the $E_{2g}^1$ and $A_{1g}$ modes, respectively, exhibit representative twofold symmetry.[42,49] The corresponding polar plots of the $E_{2g}^1$ and $A_{1g}$ intensities are depicted in Figure 1D. Our fitting results distinctly demonstrate the polar behavior of the $E_{2g}^1$ mode, with a notable 50% intensity variation between 0 and 90 degrees. The distinct polarity of the $E_{2g}^1$ peak confirms the occurrence of uniaxial strain, which in turn modulates the crystal symmetry of Fe:MoS$_2$. This symmetry alteration is known to promote field-free SOT switching by generating out-of-plane spin accumulation when current is injected along the zigzag axis (x-axis) at the Fe:MoS$_2$/Pt interface, similar to the field-free SOT switching of Fe$_3$GaTe$_2$ using WTe$_2$.[7,10]

Figure 1E depicts a hysteresis loop measurement of the Hall resistance ($R_H$) using a Fe:MoS$_2$/Pt heterostructure at 300 K. In general, the Hall resistance is expressed by

$$R_H = R_0 H + R_S M + R_{THE} \qquad (1)$$

where $R_0 H$, $R_S M$ and $R_{THE}$ denote the ordinary Hall effect (OHE), AHE and THE, respectively. When a magnetic field is applied along the out-of-plane direction (z-axis), the transverse resistance ($R_H$) exhibits sharp antisymmetric humps superimposed on a step function. The linear background with a weak positive slope at large fields arises from the OHE ($R_0 H$). The change in $R_H$ near ±0.25 T can be explained by the AHE ($R_S M$) signal, which can be approximated by a step function,[50,51] signifying excellent PMA even at room temperature.[27,30] Notably, this robust PMA arises from a monolayer of vdW material with a thickness of 0.8 nm—a highly desirable feature for applications in high-density magnetic memory.[18,20,36] The distinct antisymmetric humps close to 0 T in the Hall resistance shown in Figure 1E closely resemble those identified as THE ($R_{THE}$) in various ferromagnetic systems (discussed in Figure 4). Figure 1F shows a heterostructure with a large coercivity ($H_C$) of 3 kOe at 4 K, which monotonically decreases to 1.8 kOe at 300 K owing to the increasing thermal fluctuations in Fe:MoS$_2$.

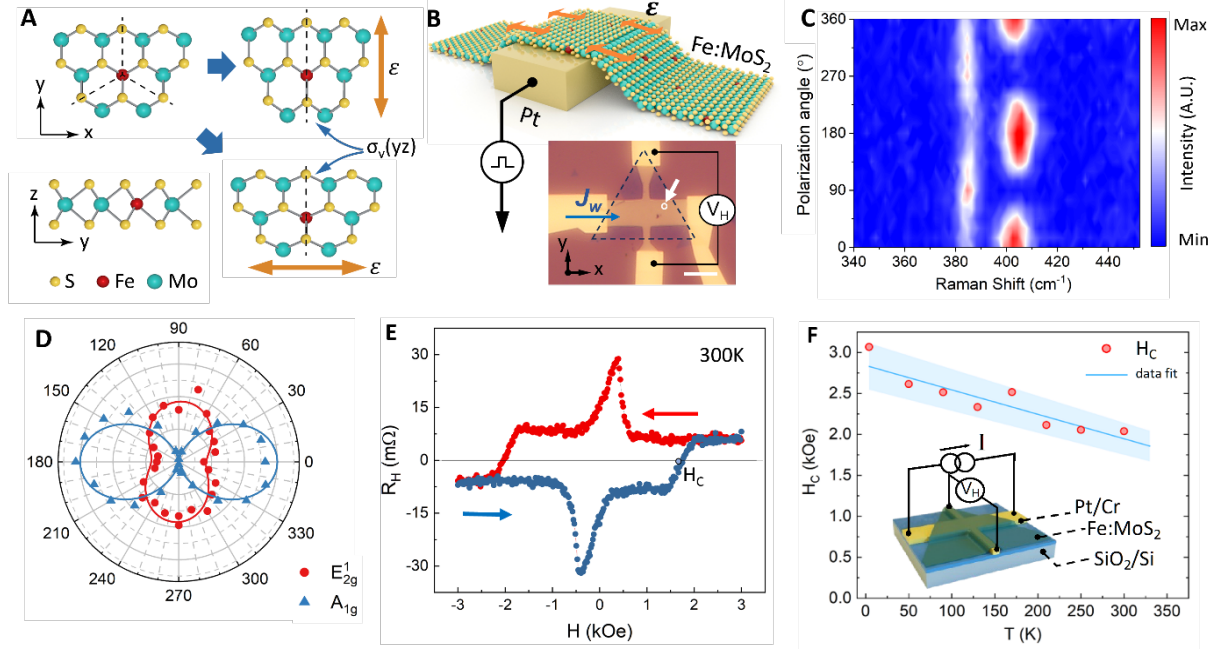

**Fig. 1. Crystal characterization: optical and Hall resistance measurements.** (**A**) Schematic illustration of the distorted lattice structure of monolayer Fe:MoS$_2$ under armchair strain. The vertical dashed line denotes the preserved mirror symmetry in the y-z plane σ$_v$ (yz) under armchair strain. (**B**) Schematic and optical images of monolayer Fe:MoS$_2$ transferred on a Pt Hall bar. The strain direction is marked as ***ε***. The optical image presents a fabricated device with a current flow direction, $J_w$, along the zigzag axis (determined through the edge of the crystal triangle), where the white arrow indicates the laser spot. (**C**) Polarized Raman spectra of the strained Fe:MoS$_2$ on a Pt electrode. (**D**) Polar plots of the $E^1_{2g}$ and $A_{1g}$ modes shown as a function of the angle between the polarizations of the incident and the scattered light. (**E**) Prototypical AHE response in the Fe:MoSe$_2$/Pt heterostructure for field sweep out of the sample plane (H ∥ z) at room temperature. (**F**) Variation in coercivity ($H_c$) obtained from AHE plots as a function of temperature.

## 2.2 Field-free, deterministic SOT switching via spin-to-spin conversion

Figure 2A illustrates the proposed SOT-driven magnetic switching mechanism within the Fe:MoS$_2$/Pt heterostructure, where the SOC and low symmetry facilitate a spin-to-spin conversion to enable perpendicular magnetization switching.[11] For the conventional field-free SOT switching of perpendicularly magnetized layers, the generation of the essential z-polarized spins is typically achieved through *charge-to-spin* conversion when a charge current flows through materials with low symmetry.[7,10,52] However, in our Fe:MoS$_2$/Pt heterostructure, the charge current mainly flows within the Pt layer due to the high resistance of Fe:MoS$_2$ (Figure

S5), thereby allowing for the generation of y-polarized spins only via the spin Hall effect (SHE). Thus, the generation of z-polarized spins needs to be facilitated by additional spin-to-spin conversions (i.e., y-polarized to z-polarized spins; see Supplementary Note 1). The linear response theory[11] predicts the conversion of an external spin current $J_{s,j}^k$ into spin accumulation $S_i$, which is encapsulated by the equation $S_i = \eta_{i,j}^k J_{s,j}^k$. Here, $\eta_{i,j}^k$ is the response tensor connecting $S_i$ and $J_{s,j}^k$, where (i, j, k) are the Cartesian coordinates (x, y, z). Specifically, i indicates the direction of the generated spin density, j denotes the direction of the spin current, and k signifies the spin polarization direction.

In accordance with the Neumann principle,[41] pristine $MoS_2$, which maintains inversion and mirror symmetry across three planes, would exhibit zero values for the response tensor components $\eta_{i,j}^k$. However, the introduction of uniaxial strain along the armchair direction (y-direction) and Fe doping breaks the inversion symmetry, in addition to decreasing the symmetry with respect to the plane $\sigma_0$ (xy) at the $Fe:MoS_2/Pt$ interface, leaving only the mirror symmetry with respect to the plane $\sigma_v$(yz) intact. This symmetry breaking allows a nonzero value of $\eta_{z,z}^y$, which results in the conversion of the y-polarized spin current $J_{s,z}^y$ generated in the Pt layer into the z-polarized spin density $S_z$ (Figure 2A, green arrow) at the $Fe:MoS_2/Pt$ interface. Assisted by this interfacial spin-to-spin conversion, an anti-damping torque $\tau_{AD}^{OOP}$ is present in our system. This unconventional, symmetry-breaking, anti-damping torque $\tau_{AD}^{OOP}$ is analogous to phenomena observed in other low-symmetry vdW materials, such as $WTe_2$ and $TaIrTe_4$.[7,26,40] The torque can be expressed as $\tau_{AD}^{OOP} \propto \hat{m} \times \hat{z} \times \hat{m}$, where $\hat{m}$ and $\hat{z}$ are unit vectors along the magnetization of $Fe:MoS_2$ and perpendicular to the interface, respectively. This torque is antisymmetric with respect to the charge current flow, $J_w$, facilitating highly efficient SOT-driven reversible magnetization switching by altering the $J_w$ direction.

To investigate current-induced SOT switching, current pulses were applied in the x-direction of the $Fe:MoS_2/Pt$ Hall bar (see Methods). As shown in Figure 2B, the magnetic state of $Fe:MoS_2$ was toggled between the z+ and z-directions without needing an external magnetic field at a notably low critical current density of $J_c = 7 \times 10^4$ A cm$^{-2}$ at 380 K. This current density corresponds to a power dissipation density of $P_{sw} = 7.5 \times 10^{12}$ W m$^{-3}$ (i.e., $P_{SW} = J_c^2 \rho$, where ρ is the resistivity of the device), which is four orders of magnitude lower than reported vdW SOT materials and three orders of magnitude lower than non-vdW SOT materials.[8,10,11,15,22]

As depicted in Figures 2B-2D, the current density required for SOT switching decreased as the temperature increased from 200 K to 380 K. This trend correlated with the diminishing coercive field ($H_c$), indicative of a reduction in effective anisotropy and the energy barrier necessary for switching,[7] as evidenced by the $H_c$ trends summarized in Figure S6. Figure 2E and Table S1 summarize the comparison chart and table of the critical current density and operation temperatures in the reported SOT switching materials. This inverse relationship between temperature and current density has also been observed in various SOT devices constructed from 2D vdW materials, including $Fe_3GeTe_2$, $Fe_3GaTe_2$ and $CrTe_2$.[6,7,19] Additional evidence supporting the reduction in the effective anisotropy and the energy barrier at elevated temperatures is a sharp transition between the -z and +z magnetic states of $Fe:MoS_2$, as depicted in Figure 2C, 2D and S7.

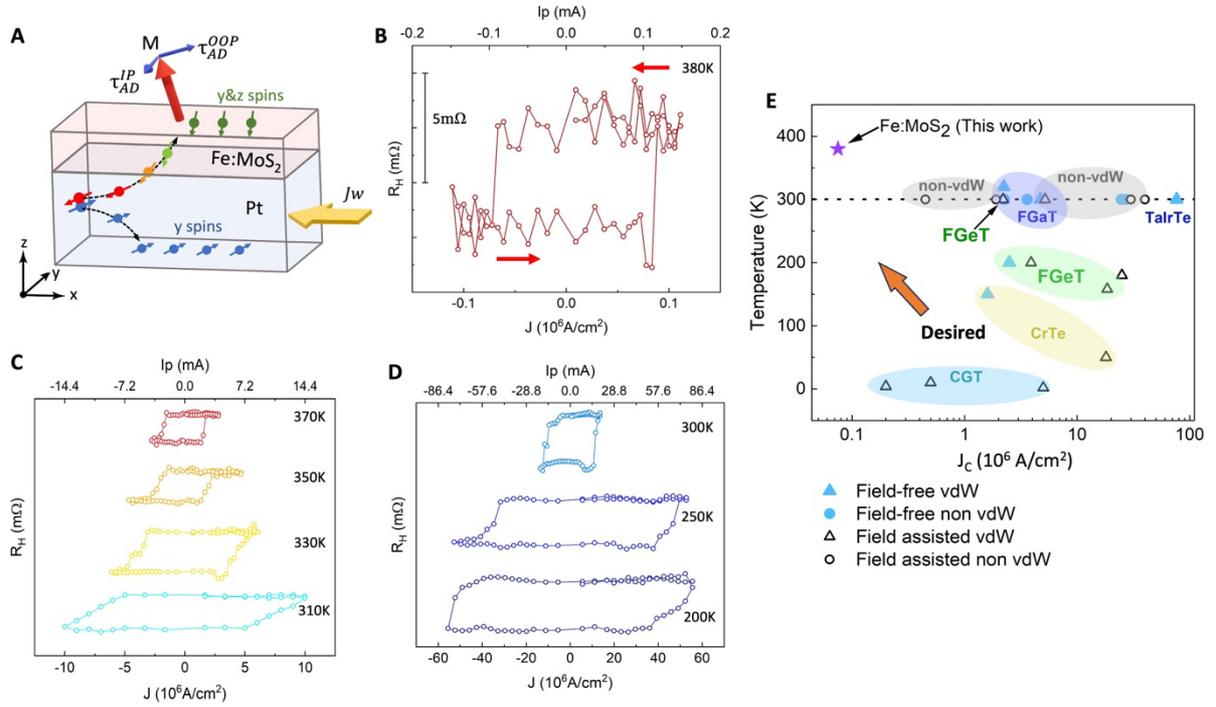

**Fig. 2. Field-free spin-orbit-torque (SOT) switching**. (**A**) Schematic illustration of the scenario in which a pulsed current is injected along the low-symmetry axis (x-axis) in the $Fe:MoS_2$/Pt heterostructure.(**B** to **D**) Field-free SOT switching of $Fe:MoS_2$/Pt at different temperatures up to 380 K. For clarity, the $R_H$ data are normalized and offset along the y-axis. (**E**) Comparison of the critical current density, $J_c$, and operation temperature for the reported SOT switching materials ($Fe_3GeTe_2$,[6,7,53–55] $Fe_3GaTe_2$,[8,10,13] $CrTe_2$,[30,56] $Cr_2Ge_2Te_6$,[57–59] $TaIrTe_4$,[26] and non vdWs[11,21,23,24,60–63]).

## 2.3 Crystal symmetry breaking-induced out-of-plane anti-damping torque

We measured the field-free switching of the perpendicular magnetization of Fe:MoS$_2$ by applying a writing current, $J_w$, along two different axes at 300 K and explored the occurrence of an out-of-plane anti-damping torque $\boldsymbol{\tau_{AD}^{OOP}}$. Monolayers of Fe:MoS$_2$ were positioned at two different angles against the Pt Hall bar, as depicted in the inset of Figure 3A (Figure 3B), to induce uniaxial strain along the zigzag (armchair) axis while retaining the mirror symmetry along the y-axis. The $J_w$ direction was set to the y-axis (armchair direction) or the x-axis (zigzag direction). As depicted in Figure 3A, the magnetization state of Fe:MoS$_2$ was first initialized to m$_z$ = +1 (with $H_z$ = +0.32T); then, $J_w$ was applied parallel to the y-axis ($\boldsymbol{J_w} \parallel y$), under which it switched to a state with near-zero Hall resistance (m$_z$, R$_H$ ≈ 0). For both pulse sweeping directions starting from m$_z$ = +1 (represented by the red and blue data points), this m$_z$ ≈ 0 state cannot be switched back to m$_z$ = +1 by reversing the current direction. This switching behavior is attributed to either demagnetization of Fe:MoS$_2$ by Joule heating or switching of the magnetization direction to the in-plane direction (x-direction) by the in-plane torque $\boldsymbol{\tau_{AD}^{IP}}$. This behavior is analogous to the demagnetized state observed in conventional bilayer heavy metal (HM)/FM SOT[12] and in the WTe$_2$/Fe$_3$GaTe$_2$ systems,[7,10] where the current was applied along the high symmetry axis.

In contrast, when a current was applied along the low-symmetry axis, as shown in Figure 3B, $\boldsymbol{\tau_{AD}^{OOP}}$ was generated, allowing the magnetization of Fe:MoS$_2$ to be toggled between m$_z$ ≈ ±0.8 upon reversing the current direction. This is consistent with magnetic switching by $\boldsymbol{\tau_{AD}^{OOP}}$, in which the magnetization direction can be reversed with a change in the writing current direction. The current density in this scenario was approximately one-tenth of the demagnetizing current density needed for $\boldsymbol{J_w} \parallel y$. Figure 3C shows the applied current pulses (1 ms pulse duration), enabling the switching of out-of-plane Fe:MoS$_2$ magnetization at room temperature.

The presence of $\boldsymbol{\tau_{AD}^{OOP}}$ in low-symmetry systems, including WTe$_2$/FM and TaIrTe$_4$/FM, was previously determined using spin-torque ferromagnetic resonance (STFMR).[26,40] It has been established that $\boldsymbol{\tau_{AD}^{OOP}}$ is not necessarily dependent on the direction of magnetization in ferromagnets but rather on the direction of the writing current $J_w$.[7,10,40] This characteristic implies that the out-of-plane magnetization (m$_z$) can be effectively modulated by applying a charge current along the x-axis, resulting in a shift in the AHE hysteresis loop. As demonstrated in Figure 3D, the AHE loop shift ($H_{SH}$) measured under current pulses shows a notable loop shift toward the positive direction when $I_x = +8$ mA, in contrast to $I_x = -8$ mA (see Methods). When repeating the measurements at different charge current magnitudes for $\boldsymbol{J_w} \parallel x$, no significant $H_{SH}$ was observed for current magnitudes smaller than 5 mA (Figures 3E and S8). This is anticipated, as $H_{SH}$ is only noticeable when $\tau_{AD}^{OOP}$ compensates for the intrinsic damping torques.[7] The asymmetry observed in the positive and negative $H_{SH}$ is likely due to the unevenness of the crystal shape (i.e., triangular shape) against the x-axis.[7,26]

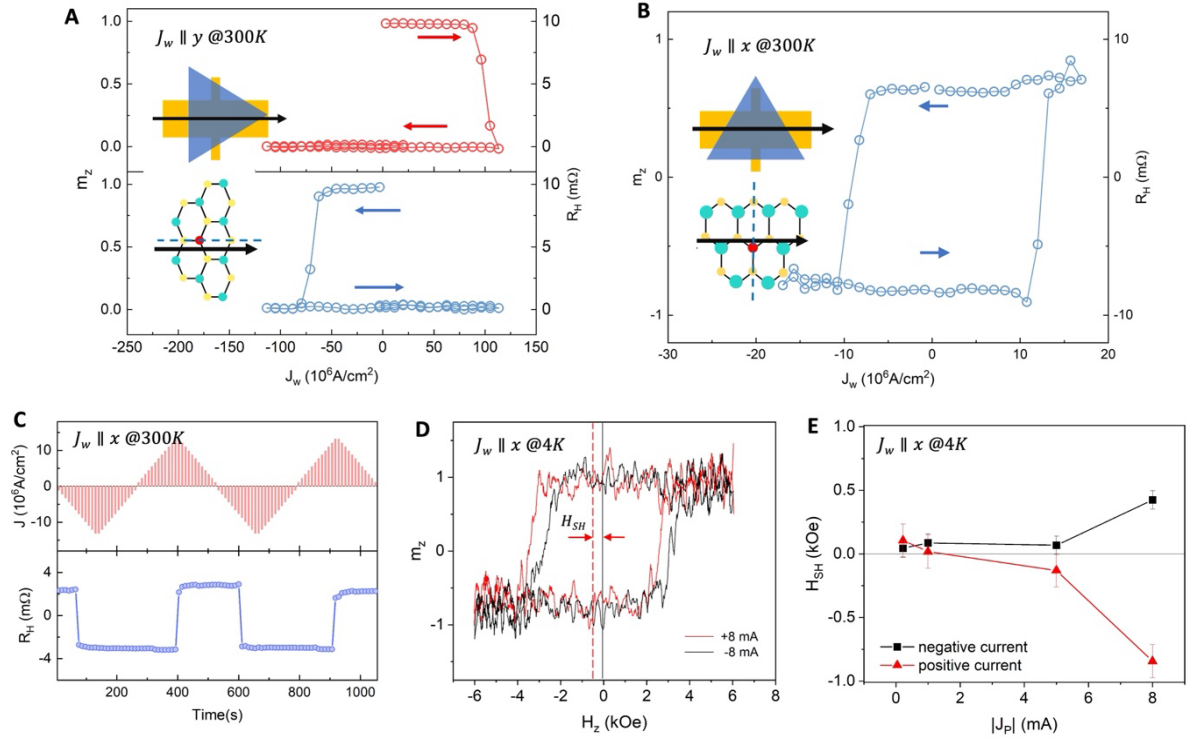

**Fig. 3. Crystal symmetry breaking in the SOT system.** (**A**) Room-temperature SOT measurements when the applied current is parallel to the y-axis (armchair direction) of the Fe:MoS$_2$ lattice. The red (and blue) curves correspond to current pulses swept from $J_w = 0 \rightarrow 120 \times 10^6$ A cm$^{-2}$ → -120 × 10$^6$ A cm$^{-2}$ → 0 (0 → -120 × 10$^6$ A cm$^{-2}$ → 120 × 10$^6$ A cm$^{-2}$ → 0). Switching from m$_z$ = +1 to m$_z$ =0 was observed for both directions, but the change did not reverse back to m$_z$ = +1 because the Fe:MoS$_2$ magnetization changed to the in-plane direction (xy plane). (**B**) Room-temperature SOT measurements when the applied current is parallel to the x-axis (zigzag direction) of the Fe:MoS$_2$ lattice. (**C**) Switching of out-of-plane Fe:MoS$_2$ magnetization upon current pulses (1 ms pulse duration) at room temperature. (**D**) Current-induced AHE loop shift owing to out-of-plane switching $\tau_{AD}^{OOP}$. The AHE hysteresis loops were measured by applying $H_Z$ at pulsed currents of 8 mA (red) and −8 mA (black). (**E**) AHE loop shift field ($H_{SH}$) as a function of the charge current amplitude |$J_p$| for both positive and negative charge currents applied along the x-axis. (inset of a, b: Schematics of the Fe:MoS$_2$ single crystal (blue), the Hall bar (yellow), and the strained crystal lattices.)

## 2.4 Field- and temperature-dependence of THE

Figure 4A shows the temperature dependence of the Hall resistance obtained by sweeping the magnetic field. The antisymmetric humps observed near zero in the Hall resistance were preserved from 4 K to 300 K without a significant shape change. The humps are similar to those reported for THE in a range of ferromagnetic systems, which possess ferromagnetic order arising from intrinsic origins or doping.[51,64] A recurring observation in the literature is that the hump location typically coincides with the AHE domain reversal, which can be attributed to either THE[50,51,64–66] or two-channel AHE.[67,68] Accordingly, we further decomposed the Hall signal (red curve) into two channels with humps near the domain reversal. As shown in **Figure 4.B,** $R_{H1}$ (blue dashed line) presents a Hall loop with domain reversal near 'hump a' at $\pm2.5$ kOe, while the remaining Hall signal, $R_{H2}$ (black dashed curve), shows a Hall loop with domain reversal near 'hump b' at $\pm0.5$ kOe.

THE signal is known to be closely related to the Dzyaloshinskii-Moriya interaction (DMI)-induced magnetic skyrmions, where moving spin-polarized electrons acquire the skyrmion-induced Berry phase by adjusting their spins to the local spins of the skyrmion texture, giving rise to the observed humps.[69] The presence of an appropriate DMI intensity ($D$) within an FM/HM system is known to be crucial in facilitating field-free SOT magnetization switching.[25,70] This switching process is initiated by domain nucleation at one edge of the FM, followed by domain wall propagation toward the opposite edge, as proposed by other FM/Pt-based examples and micromagnetic simulations.[25] We have also validated that the formation of THE originates from the interfacial DMI by replacing the Pt layer with $Bi_2SeTe_2$, a topological insulator film. In the $Fe:MoS_2/Bi_2SeTe_2$ heterostructure, as illustrated in Figure S9, the AHE is observed in the absence of any characteristic THE hump, providing a clear distinction between the AHE and THE and reinforcing the role of the interfacial DMI manifested by the THE. To further exam if those two humps are from other effect such as spateial inhomogenility,[71] interface modification[68] or strain, [67,72] we studied the possible interpretations of "hump a" and "hump b" in Figure 4B (see supplymenray text 3). The origin of 'hump a' suggests the presence of THE, but further spatial scanning of the magnetic vector is required to fully understand the origin of 'hump b'. Figure 4C shows the temperature and magnetic field dependence of humps, consolidating the robust THE response from "hump a" across the entire temperature range. TFigure 4D shows the extracted peak positions of the "hump a" ($H_a$) that monotonically decrease as the temperature increases above 170 K. Whereas similar humps can be produced by alternative methods, such as competing ferromagnetic phases caused by interface-induced AHE sign reversal[68] or nonhomogeneous samples,[73] magnetic circular dichroism (MCD) analysis revealed no visible magnetic subdomains in the Fe:MoS$_2$ samples.[37] Furthermore, the photoluminescence (PL) mapping of Fe:MoS$_2$ corroborates the uniformity of the defect luminescence and Fe-related peak (Figure S10) to rule out the above scenarios. Another possibility is that the humps in AHE measurements can manifest as interface-induced sign reversal, typically requires two ferromagnetic layer with composition difference.[68] However, such a phenomenon is not possible in Fe:MoS$_2$/Pt heterostructure, as we only use a monolayer Fe:MoS$_2$ for Hall measurement.

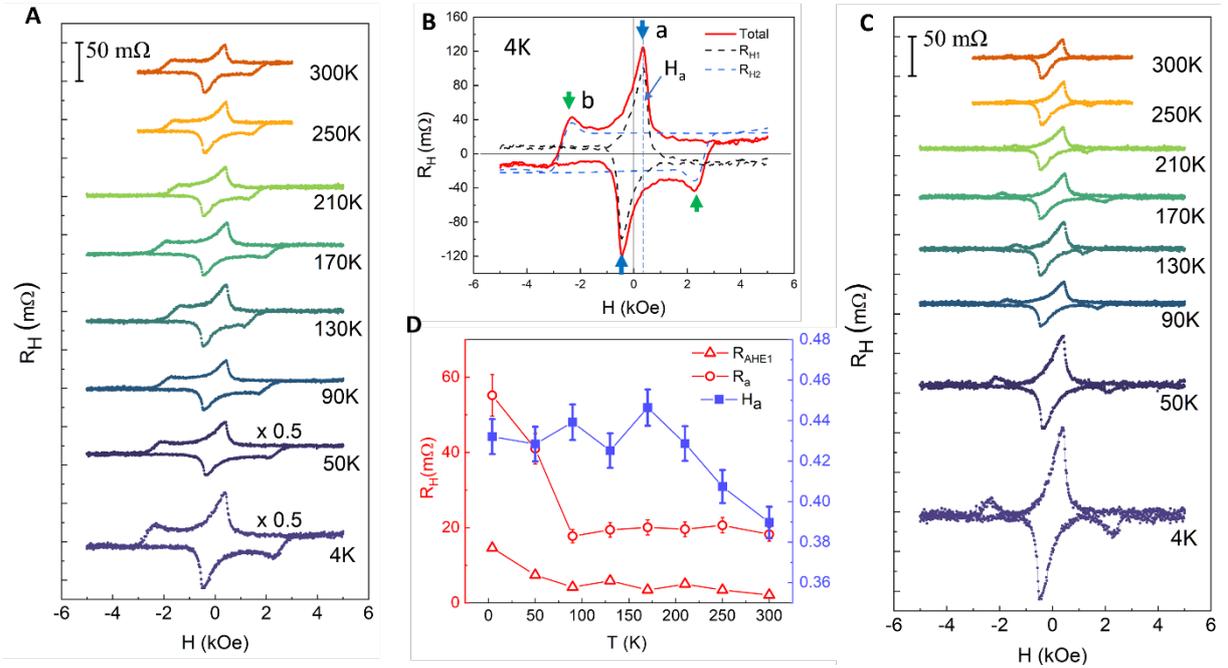

**Fig. 4. Topological Hall effect (THE) in the Fe:MoS$_2$/Pt heterostructure.** (**A**) Measured AHE loops from 4 K to 300 K. (**B**) Hall resistance after subtracting the ordinary Hall effect. The red curve was further decomposed into two channels ($R_{H1}$ and $R_{H2}$), with each channel exhibiting humps near the magnetic domain reversal. The blue arrows indicate "hump a" near ±2~2.5 kOe, observed before the domain reversal at ±3 kOe, while the green arrows mark "hump b" near ±0.5 kOe. (**C**) Plot of the temperature and applied field of the deduced hump responses. The white arrows indicate the direction of the field scanning. (**D**) Temperature dependence of the signals from $R_{H1}$. With Anomalous Hall resistivity from $R_{H1}$ (red triangle) and topological Hall resistivity (red circle). The positions of "hump a" maximum $H_T$ are shown in blue.

## 3. Conclusion and Outlook

We have demonstrated ultralow-power, field-free, deterministic, and nonvolatile perpendicular magnetization switching by SOT up to 380 K using a DMS, monolayer Fe:MoS$_2$ through interfacial coupling with a Pt Hall bar. The induced strain in the crystal caused a reduction in the crystal symmetry in monolayer Fe:MoS$_2$. We showed the anisotropy of field-free SOT on the crystallographic axis of Fe:MoS$_2$ by injecting current in the armchair or zigzag direction. The perpendicular magnetic switching of Fe:MoS$_2$ was confirmed by measuring the AHE in the adjacent Pt layer. An apparent AHE loop shift was observed at a zero in-plane magnetic field, verifying the existence of $\tau_{AD}^{OOP}$ through spin-to-spin conversion in the Fe:MoS$_2$/Pt heterostructure with reduced crystal symmetry. We demonstrated a switching current density of $7 \times 10^4$ A/cm$^2$, nearly two orders of magnitude lower than reported field-free SOT systems. The power dissipation density of $P_{sw} = 7.5 \times 10^{12}$ W m$^{-3}$ is four orders of magnitude lower than reported vdW SOT materials and three orders of magnitude lower than non-vdW SOT materials.[6–10,15,22] The magnitude of the topological Hall resistance persists up to 300 K, indicating that a robust magnetic skyrmion is maintained over a wide temperature range and that the topological features at the Fe:MoS$_2$/Pt interface are stable. This strong manifestation of THE was attributed to the interfacial DMI, which contributed to field-free SOT switching in the Fe:MoS$_2$/Pt heterostructure. This SOT application using a 2D monolayer DMS provides a new pathway for developing highly power-efficient spintronic devices.

## 4. Experimental Section

*Synthesis of MoS2 monolayers and Fe-doped MoS2:*
MoS$_2$ monolayers were synthesized via low-pressure chemical vapor deposition (LPCVD). For growth, a thin layer of MoO$_3$ was deposited via Joule heating thermal evaporation onto a Si substrate covered with a 300 nm thick thermal oxide layer. Another SiO$_2$/Si substrate was in direct contact with the MoO$_3$-deposited substrate during growth.

To substitutionally dope Fe into the MoS$_2$ monolayers, Fe$_3$O$_4$ particles were applied to the SiO$_2$/Si substrate before contacting the MoO$_3$-deposited substrate. For a uniform distribution of Fe$_3$O$_4$ particles, the substrate was washed with deionized (DI) water, creating a thin layer of water on the SiO$_2$ surface. Subsequently, the Fe$_3$O$_4$ particles were applied onto the surface of the substrate. The substrate was then annealed at 110 °C for 5 minutes on a hot plate. The furnace was heated at a ramp rate of 18 °C/min and held at 850 °C for 15 minutes. Ar gas was initially supplied at a rate of 30 sccm during heating until the temperature reached 300 °C. Then, $H_2$ gas was introduced at a rate of 15 sccm as the temperature increased to 760 °C. Sulfur was added to the furnace once the temperature reached 790 °C. The temperature and gas flow were controlled throughout the procedure.

*Fabrication of Fe:MoS2/Pt heterostructures:*

The four Hall bar contacts (Pt/Cr - 8/5 nm) with dimensions of 2 mm × 5 mm were patterned on a $SiO_2$/Si substrate via e-beam lithography. First, the substrate was spin-coated with PMMA and then baked at 180 °C for 90 seconds. Subsequently, the photoresist was exposed to the e-beam and developed using a 3:1 mixture of 4-methyl-2-pentanone (MIBK) and 2-propanol (IPA). The Pt/Cr layers were deposited onto the patterned photoresist layer via e-beam evaporation. The photoresist was removed through a lift-off process involving rinsing with acetone and IPA to define Pt/Cr electrodes.

Fe:$MoS_2$ monolayers grown on $SiO_2$ were coated with a thin layer of PMMA (950 A4) using a dropper. The chip was left under ambient conditions to dry for 60 min. The chip was floated in 10% KOH (aq) for 10 min, after which the Si substrate sunk 10% KOH (aq) due to $SiO_2$ etching, leaving the PMMA/$MoS_2$ floating on the surface of the KOH solution. Next, the PMMA/$MoS_2$ was cleaned in DI water, scooped using another substrate (i.e., target substrate), and dried under ambient conditions for 1 hour. A PMMA film with a Fe:$MoS_2$ crystal was attached beneath poly(dimethylsiloxane) (PDMS) and stamped with Fe:$MoS_2$ aligned atop the Pt electrode. The PMMA was removed using warm acetone, followed by an IPA rinse. The fabricated Fe:$MoS_2$/Pt sample was loaded in a furnace (MTI OTF 1200X) with a 2" quartz tube and annealed at 120 °C for 20 min to enhance adhesion, followed by soaking in 2-hour chloroform ($CHCl_3$) at room temperature to remove organic residues thoroughly.

*Hall resistance measurement:*

Magnetoresistance measurements were conducted using a variable temperature insert (VTI) within an attoDRY 1100 cryostat, with a temperature range from 4 K to 300 K and a vertical magnetic field of up to 9 T at 4 K. The Hall resistance was measured using a four-probe by applying 0.1-90 mA of d.c. current with a Keithley 2636B source meter and measuring the voltage drop ($V_{xy}$) with a Keithley 2182A nanovoltmeter. The AHE loop shift was measured with the same setup by applying a 10 ms pulse using an SDG 2122x (Siglent Technologies) waveform generator. The loop shift measurement was conducted only at 4 K to avoid quenching of the superconducting magnet in an attoDRY 1100.

*SOT measurements:*

The field-free SOT switching measurements were conducted within an Atto-DRY 1100 cryostat capable of applying a magnetic field of up to 0.5 T at 300 K. Pulses were generated using a Keithley 2636B source meter, with each pulse duration of 1 ms. To mitigate noise, the SOT device was allowed to rest for 5 s after each pulse cycle. The $R_H$ values were determined by averaging the Hall resistivity values for 5-10 s, measured with a d.c. of 50-90 µA. For SOT data measurements above 300 K, the same setup was employed in a Cryostat ST-500 (Lakeshore Cryotronics) instrument equipped with a custom-built magnet capable of applying 0.35 T in the z-direction. The SOT data from the Atto-DRY 1100 and Cryostat ST-500 for temperatures between 200 K and 300 K were compatible.

**Supporting Information**

Supporting Information is available from the Wiley Online Library or from the author.

**Acknowledgments**

E.H.Y. acknowledges financial support from the AFOSR under Grant FA9550-11-10272 and the NSF under Grant ECCS-1104870. S. S. acknowledges financial support from the NSF under Grants NSF-DMR-1809235, NSF-EFRI-1641094, and ECCS-MRI-1531237. D.S. acknowledges financial support from the DoE under award number DE-SC0020992. This research used microscopy resources, partially funded by the NSF under Grant NSF-DMR-0922522, within the Laboratory for Multiscale Imaging (LMSI) at Stevens Institute of Technology. This work was also partially carried out at the Micro Device Laboratory (MDL) at Stevens Institute of Technology, funded with support from W15QKN-05-D-0011. The authors thank Yuxing Liu and Shivani Bhawsar for their assistance in manuscript preparation.

**Author contributions**

S.C. and E.H.Y. conceived the experiments for synthesizing and characterizing the materials and devices. S.C., M.F., and Y.Z. carried out the synthesis of Fe:MoS$_2$. S.C., M.F., Z.T. and L.X carried out optical, Raman, PL, SEM, and XPS characterization and analysis. M.F. and A.S.S. performed AFM characterization. R.S. and S.C. designed the SOT device. S.C. and M.F. performed the fabrication, with contributions from X.Z. and Y.Z.. S.C. designed the electromagnetic measurements, and S.C. and A.S.S. built the experimental setup. N.L. made an initial contribution to the measurement setup. S.C. and Z.T. characterized SOT systems. S.C., R.S., S.S., D.S., and E.H.Y. conducted data analysis. All authors discussed the results and contributed to the final manuscript.

Supporting Information

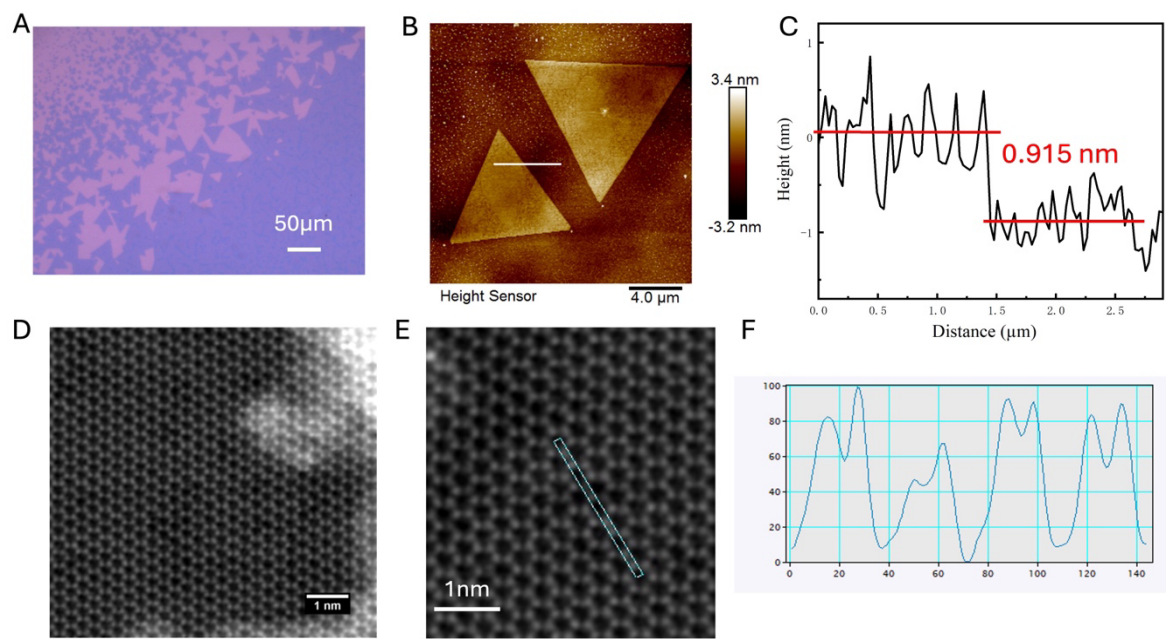

**Fig. S1. Characterization of CVD-grown monolayer Fe:MoS$_2$** (**A**) Optical image of CVD-grown Fe:MoS$_2$ monolayers. (**B** and **C**) Atomic force microscopy (AFM) image confirming the monolayer thickness of Fe:MoS$_2$. (**D** and **E**) High-angle annular dark-field scanning transmission electron microscopy (HAADF-STEM) characterizing the Fe doping. (**F**) Intensity spectra from the selected area shown in (**e**) indicating that the Fe atom exhibits a 40% lower intensity than Mo.

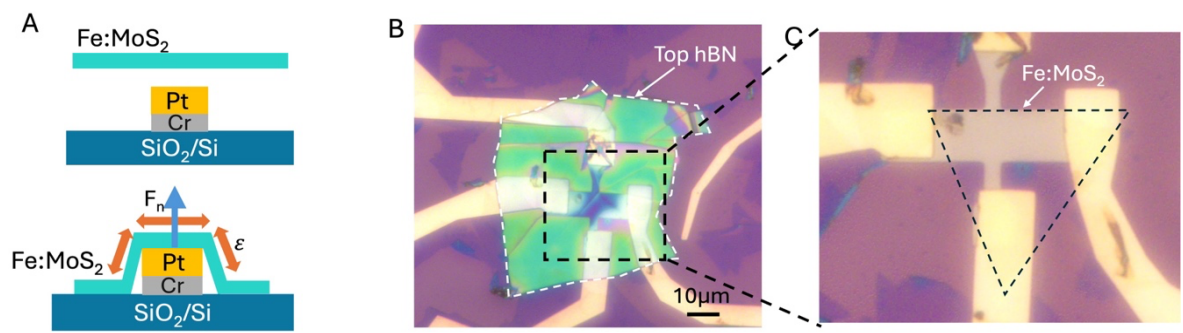

**Fig. S2. Typical Fe:MoS$_2$/Pt heterostructure.** (**A**) Crossectional schematic of the Fe:MoS$_2$/Pt geometry, inducing a uniaxial strain in Fe:MoS$_2$ perpendicular to the Pt Hall bar edge. (Not to scale) (**B**) Optical image of a typical sample, showing a 20 nm-thick hBN atop Fe:MoS$_2$/Pt heterostructure. (**C**) Zoom-in view of the area indicated by the black dashed square in (**B**).

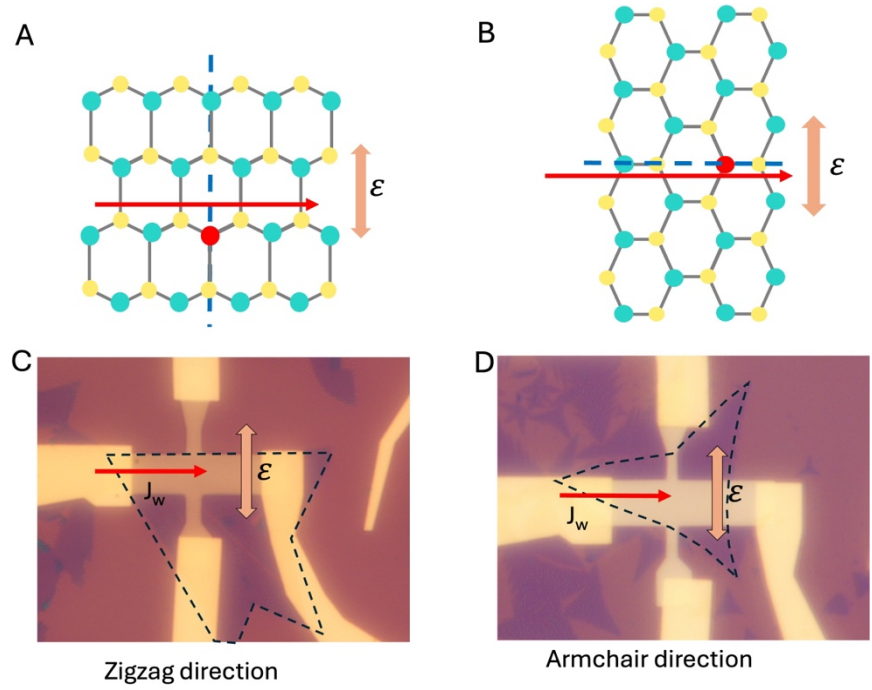

**Fig. S3. Fe:MoS₂ crystals with different angles relative to the Pt Hall bar.** (**A** and **B**). Schematic of Fe:MoS$_2$ lattices with uniaxial strain directions altering the crystal symmetry. Under both configurations, the mirror symmetry is maintained along the armchair direction of the crystal. (**C** and **D**). Optical images of Fe:MoS$_2$ crystals transferred on the Pt Hall bar, showing current flows parallel to the zigzag (**C**) and armchair (**D**) directions, respectively.

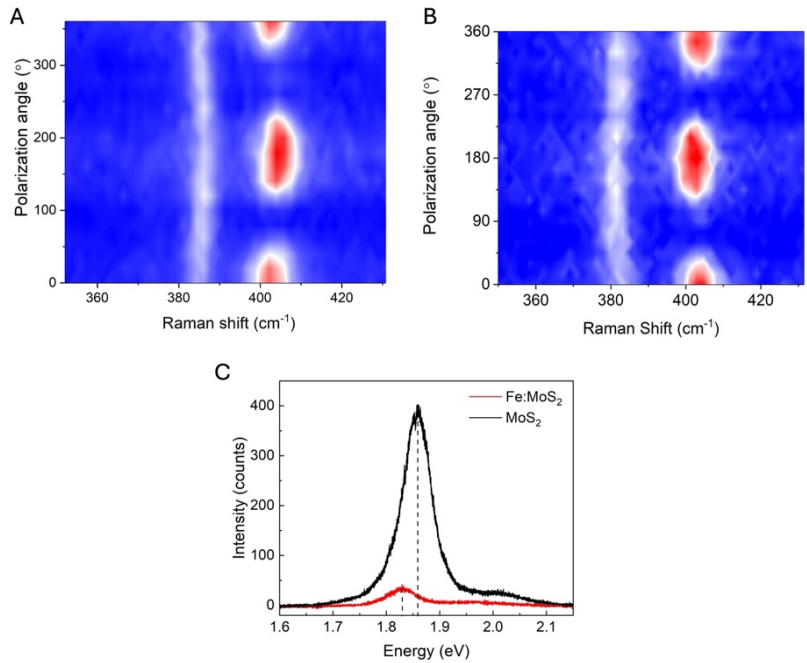

**Fig. S4. Polarized Raman and photoluminescence spectra of monolayer Fe:MoS$_2$ and MoS$_2$ on SiO$_2$.** (**A** and **B**) Polarized Raman spectra revealing that the $E^1_{2g}$ mode, at approximately 383 cm$^{-1}$, exhibits no polarization from both monolayer Fe:MoS$_2$ and MoS$_2$ on SiO$_2$. This suggests that Fe doping does not alter the crystal symmetry. (**C**) Photoluminescence (PL) spectra of Fe:MoS$_2$ and MoS$_2$ on SiO$_2$, respectively, illustrating a redshift and quenching effect due to Fe doping.

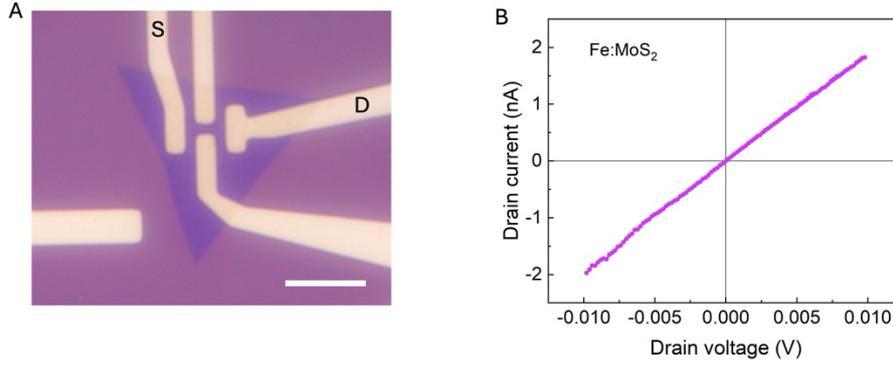

**Figure S5: Fe:MoS$_2$ conductivity a.** Optical microscope image of a monolayer single crystal Fe:MoS$_2$ on Au electrodes. Scale bar, 10$\mu$m. **b.** Source-drian IV characteristics of a monolayer Fe:MoS$_2$ at 300K.

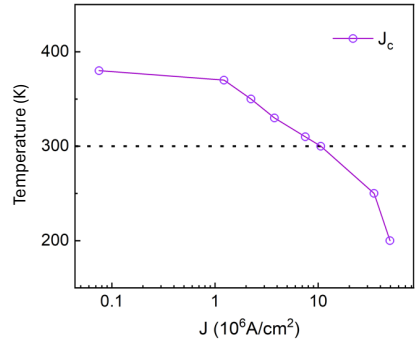

**Figure S6. Temperature-dependent critical current density for SOT Switching.** The critical current density decreases as temperature increases, owing to the reduced energy barrier for switching at elevated temperatures.

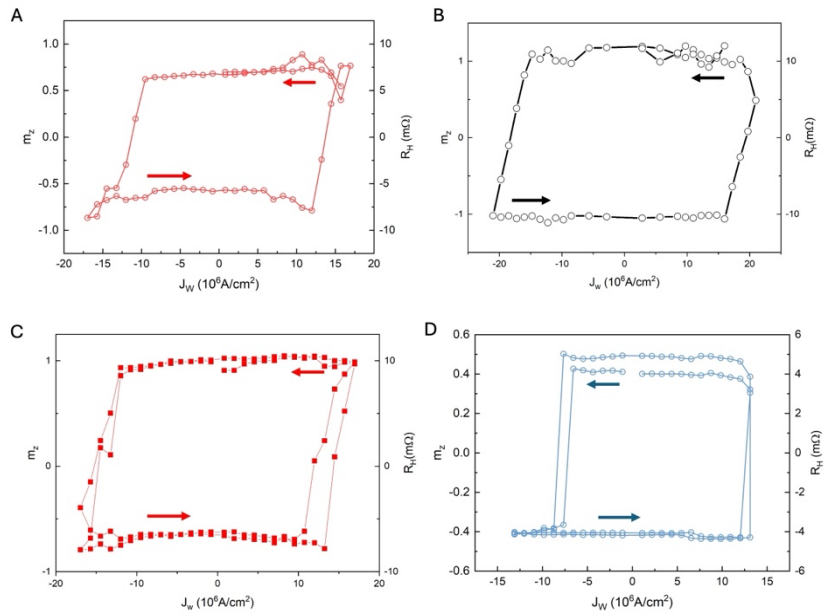

**Fig. S7. Field-free SOT in different devices. (A to D)** Demonstration of field-free SOT switching in four distinct devices by applying 2 ms current pulses along the x-axis ($J_w \parallel x$) at 300K. This showcases the reproducibility and reliability of the SOT device.

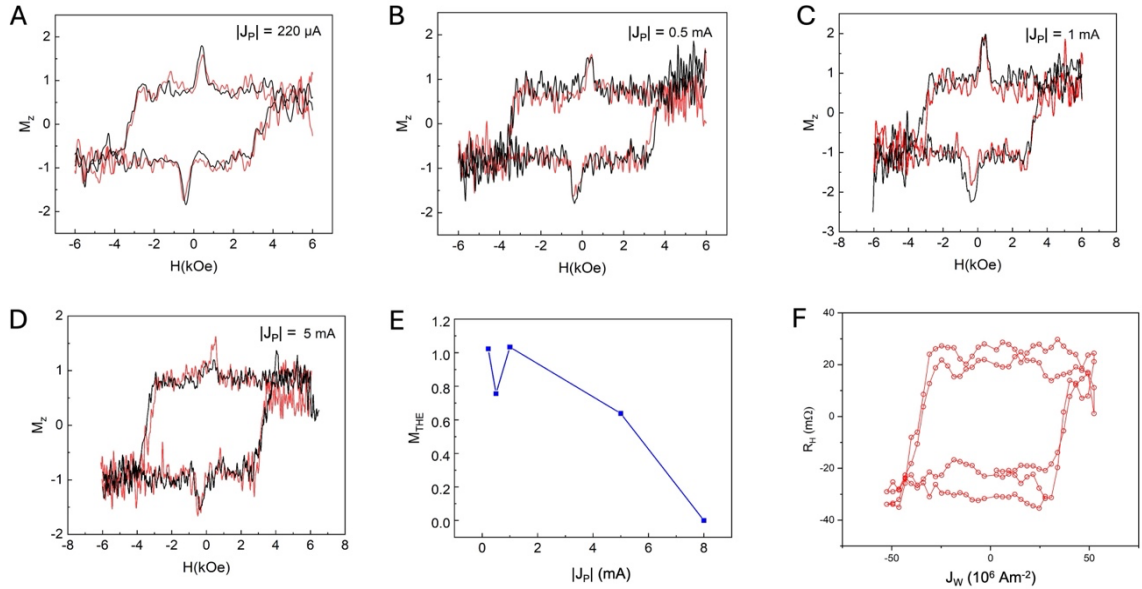

**Fig. S8. AHE loop and field-free SOT at 4 K (A to D)** Depiction of the AHE loop at 4 K with pulsed currents under the external magnetic field (*H*) perpendicular to the sample plane. The pulse duration was 10 ms with the current amplitude ranging from 0.22 mA to 5 mA. The Hall resistance is normalized as $M_Z = R/R_{AHE}$. (**E**) The amplitude of THE signal ($M_{THE} = R_{THE}/R_{AHE}$) decreases monotonically with increasing $|J_P|$. (**F**) Field-free SOT switching at 4 K.

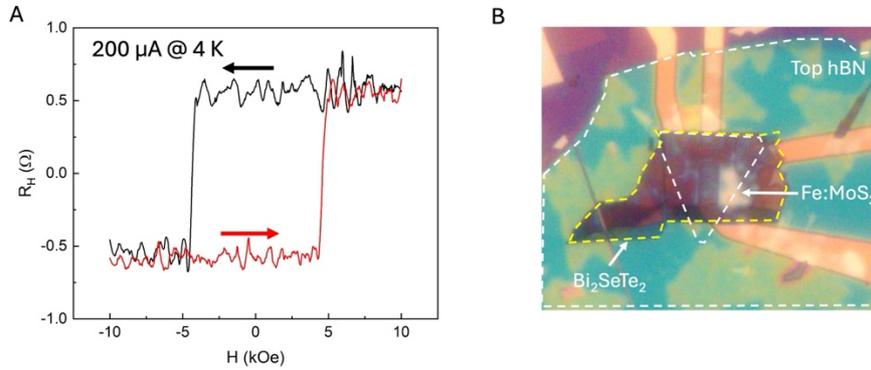

**Fig. S9. Anomalous Hall effect in Fe:MoS$_2$/Bi$_2$SeTe$_2$** (**A**) AHE measurement of a Fe:MoS$_2$/Bi$_2$SeTe$_2$ heterostructure at 4 K with an external magnetic field applied out-of-plane (z-direction). (**B**) Optical image of the Fe:MoS$_2$/Bi$_2$SeTe$_2$ heterostructure, with layers marked. The thickness of the Bi$_2$SeTe$_2$ layer is 23 nm.

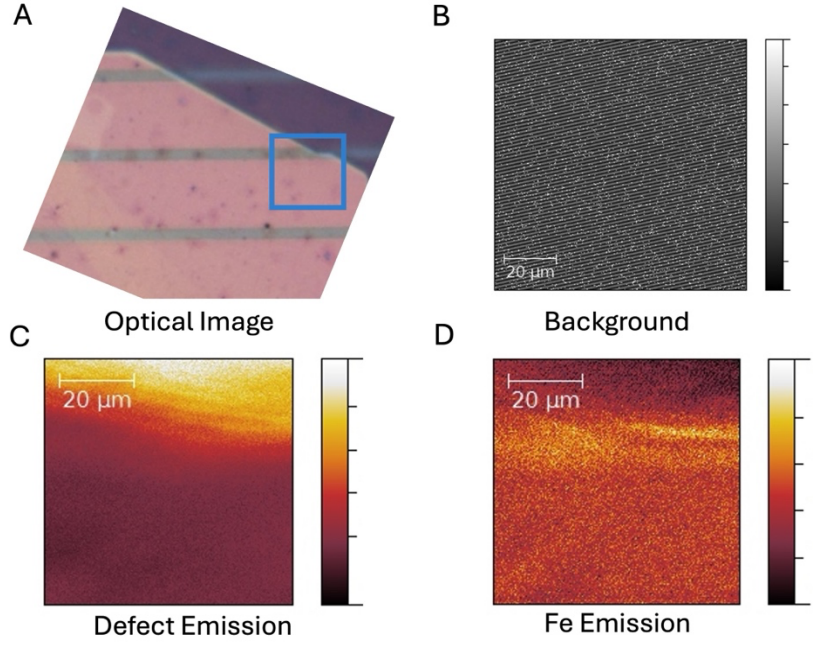

**Fig. S10. PL mapping of a typical monolayer Fe:MoS$_2$ crystal performed at 4 K.** (**A**) Optical image of a monolayer Fe:MoS$_2$ transferred onto a hBN layer. (**B**) Hyperspectral PL background signal when scanning off-sample. (**C**) Mapping of defect emission, filtered within the 650 nm to 800 nm range. (**D**) Fe-related emission, filtered within the 540 nm to 560 nm range.

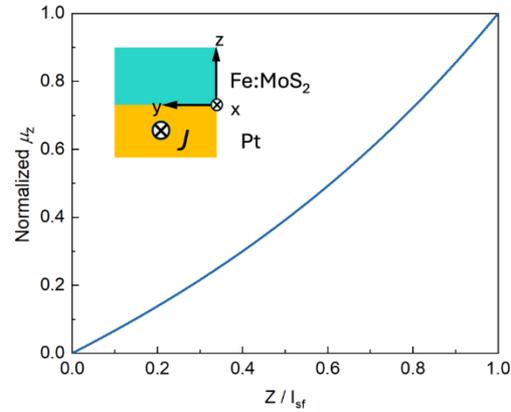

**Figure S11**. Normalized z-polarized spin potential ($\mu_z(z)$) as described by Equation (9) for 0< $z/l_{sf}$<1. The inset shows the coordinate system of the device, where z=0 represents to the Fe:MoS$_2$/Pt interface. The charge current (*J*) is injected along the x-axis.

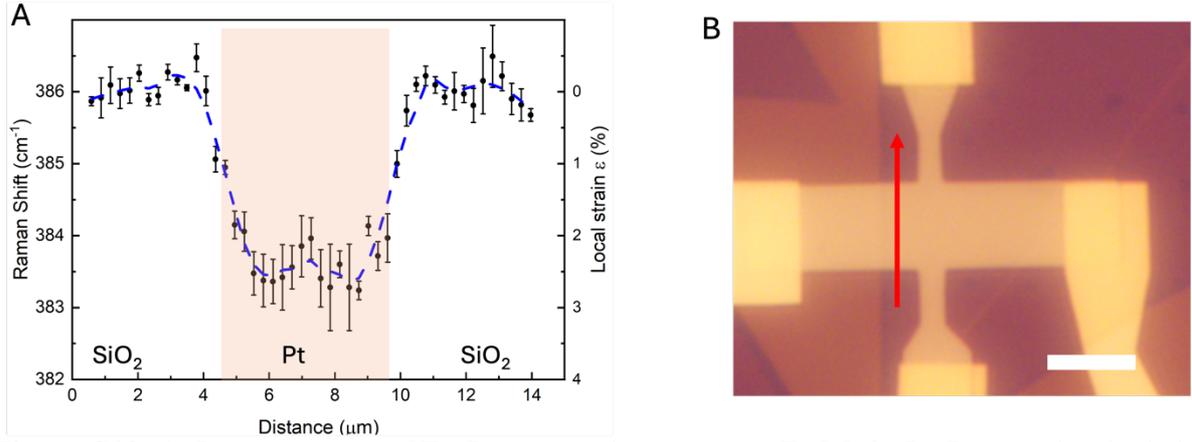

**Figure S12**. **A**. Line scanning of $E_{2g}$ Raman peak position in Fe:MoS$_2$/Pt. Data in the shaded area was obtained when the Raman laser spot was located inside the Pt Hall bar area. The error bars were determined by s.d. of each measurement. **B.** Optical image of the Fe:MoS$_2$/Pt sample. The red arrow shows the Raman scanning direction.

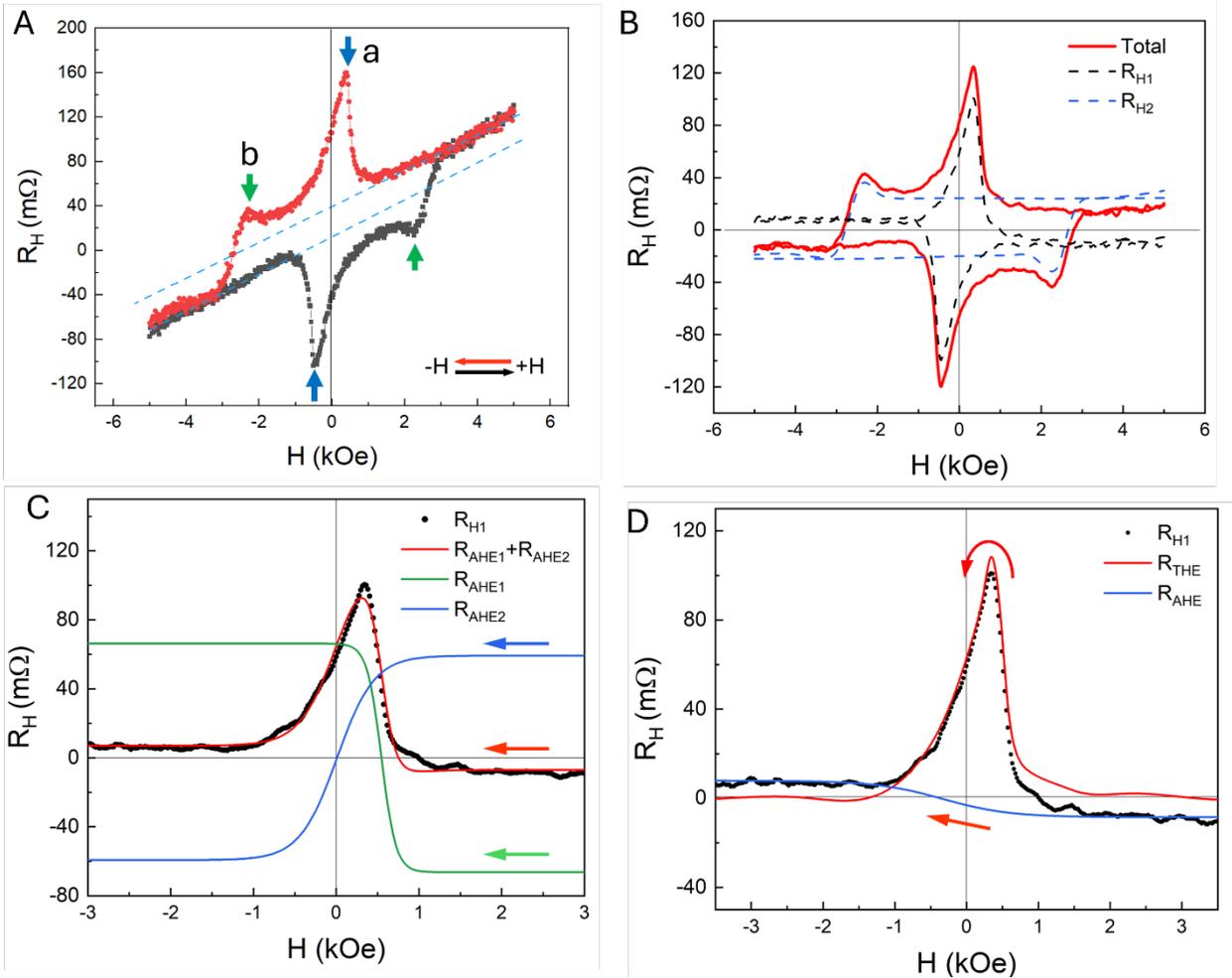

**Figure S13**. **A**. Hall resistance data ($R_H$) in Fe:MoS$_2$ monolayers at 4K with $I_{xx}$ =140 μA. The blue arrows indicate "hump a" near ±2~2.5 kOe, observed before the domain reversal at ±3 kOe, while the green arrows mark "hump b" near ±0.5 kOe. The light-blue dashed lines represent the linear background from the ordinary Hall effect. **B**. Hall resistance after subtracting the ordinary Hall effect. The red curve was further decomposed into two channels ($R_{H1}$ and $R_{H2}$), with each channel exhibiting humps near the magnetic domain reversal. **C**.

Fitting for the upper half of $R_{H1}$ with two tanh functions (two AHE channels) as represented by the blue and green curves. All curves scanned from 3 kOe to -3 kOe. **D.** Fitting for the upper half of $R_{H1}$ using one tanh function, indicated by a solid blue curve, with ferromagnetic domain reversal near -0.5 kOe. The remaining signal is attributed to THE. The red arrow on the bottom indicates the scan direction.

**Table S1.**

**Table S1 Comparison of SOT current density and operating temperature**

| | Materials | Maximum temperature of SOT switching (K) | Switching current density ($10^6$ A cm$^{-2}$) | $H_X$ (Oe) |
|---|---|---|---|---|
| | Py/TaIrTe$_4$ [26] | 300 | 76.5 | 0 (Field Free) |
| | L1$_0$-Fe$_{45}$Cr$_5$Pt$_{50}$ [24] | 300 | 27 | 0 (Field Free) |
| | CuPt/CoPt [23] | 300 | 24.7 | 0 (Field Free) |
| | [Pt$_{0.6}$/Hf$_{0.2}$]$_6$/CoFeB [62] | 300 | 3.6 | 0 (Field Free) |
| | BPBO/LSMO [21] | 300 | 0.4 | 100 |
| | Ta/CoFeB [61] | 300 | 0.4 | 1000 |
| | W/CoFeB [63] | 300 | 0.3 | 1 |
| | Pt/CoFeB [60] | 300 | 0.3 | 4 |
| vdW magnets | (Bi$_{1-X}$Sb$_X$)$_2$/Cr$_2$Ge$_2$Te$_6$ [59] | 2 | 5 | 1000 |
| | Cr$_2$Ge$_2$Te$_6$/Ta [58] | 4 | 0.2 | 1000 |
| | Cr$_2$Ge$_2$Te$_6$/Ta [57] | 10 | 0.5 | 200 |
| | CrTe$_2$/ZrTe$_2$ [30] | 50 | 18 | 200 |
| | Fe$_3$GeTe$_2$/Pt [54] | 180 | 25 | 3000 |
| | WTe$_2$/Fe$_3$GeTe$_2$ [55] | *200* | *3.9* | *300* |
| | WTe$_2$/Fe$_3$GeTe$_2$ [7] | 200 | 2.5 | 0 (Field Free) |
| | Fe$_3$GaTe$_2$/Pt [13] | 300 | 5.2 | 3000 |
| | Fe$_3$GaTe$_2$/Pt [8] | 300 | 4.7 | 0 (Field Free) |
| | WTe$_2$ /PtTe$_2$ /CoFeB [11] | 300 | 2.25 | 0 (Field Free) |
| | Bi$_2$Te$_3$/Fe$_3$GeTe [6] | 300 | 2.2 | 2000 |
| | WTe$_2$/Fe$_3$GaTe$_2$ [10] | 320 | 2.23 | 0 (Field Free) |
| | ***Fe:MoS$_2$/Pt (Our work)*** | **380** | **0.075** | **0 (Field Free)** |

**Supplementary Text**

1. Spin-to-spin conversion

1.1 Spin current injection

The accumulation of y-polarized spins gives rise to a spin chemical potential ($\mu_s$), resulting in spins being injected into Fe:MoS$_2$, which can be expressed by :

$$j_{Ps,z} = -\frac{\sigma_P}{2e}\partial_z\mu_s + \theta_p\sigma_P E_x y \qquad (1)$$

where $\sigma_P$ and $\theta_p$ are the electrical conductivity and the spin Hall angle of Pt, respectively. $E_x$ is the applied electric field along the x direction. The spin-to-spin conversion occurs primarily in the z-direction, given the monolayer thickness of Fe:MoS$_2$ (approximately 0.8 nm, Figure S1c). The resulting spin current in the z-direction is given by:

$$j_{Fs,z} = -\frac{\sigma_F}{2e}\partial_z\mu_s + \sigma_F E_x z \qquad (2)$$

where $\sigma_F$ is the Hall conductivity of Fe:MoS$_2$. With the spin current $j_{Fs,z}$ contributing by both the in-plane electric field along the x direction ($E_x$), and the spin diffusion pushed by $\mu_s$.

This spin-to-spin conversion process can be described using linear response theory, as presented by Wang et al.[11] From the quantum Liouville equation, the nonequilibrium spin density (δS) can be expressed by modeling the symmetry breaking in the x-z plane as a Rashba-like spin orbit coupling (SOC):

$$\delta S = \sum_k [\delta s_k^0 + 2\alpha_R \tau_s (k \times y) \times \delta s_k^0] \qquad (3)$$

where $\delta s_k^0$ refers to the resolved contribution in the crystal momentum space of electrons, $k$ is the crystal momentum of electrons, $\alpha_R$ is the Rashba parameter up to the first order, and $\tau_s$ is the spin relaxation time. For a qualitative explanation of the expression of spin current and spin flipping time, we employed a continuity equation within spin dynamics. The spin current ($\mathcal{J}_k$) and spin density ($S_k$) are related as follows[74]:

$$\frac{\partial S_k}{\partial t} + \nabla \cdot \mathcal{J}_k = -\frac{S_k}{\tau_s} \qquad (4)$$

where $k$ represents the spin direction ($k$ = x,y,z). As a result of the spin Hall effect (SHE), the spin current has the form:

$$\mathcal{J}_k = -D_s \nabla S_k + \sigma_\theta E_i \qquad (5)$$

where $\sigma_\theta, E_i$, and $D_s$ represent the spin-hall conductivity, electric field along the $i$ direction and spin diffusion coefficient, respectively. In a system with SOC, the spin density is related to the spin chemical potential ($S_k \propto \mu_k$)., leading to the following continuity equation:

$$\frac{\partial \mu_k}{\partial t} + \sum_i \partial_i J_{s,i}^k = -\frac{\mu_k}{\tau_s} \qquad (6)$$

where $J_{s,i}^k$ represents the external spin current with $k$-spin, flowing in $i$-direction. When the system reaches a steady state, we have $\frac{\partial \mu_k}{\partial t} = 0$. With the y-spins being converted into z-spins in Fe:MoS$_2$ ($J_{Fs,i}^z$), we incorporate the spin accumulation: $S_z = \eta_{y,z}^z J_{s,y}^z$ into quation (4) for z-spins which becomes:

$$-\frac{\mu_z}{\tau_{sf}} \sim \partial_z J_{Fs,z}^z + \eta_{y,z}^z J_{Ps,y}^z \qquad (7)$$

where $\tau_{sf}$ and $\eta_{y,z}^z$ are the spin-flipping time and response tensor for converting y-spin to z-spin in Fe:MoS$_2$, respectively. The $J_{Fs,z}^z$ and $J_{Ps,y}^z$ are z-spins from Fe:MoS$_2$ and y-spins from Pt, respectively, both flowing in the z-direction.

1.2 Spin chemical potential

Next, to qualitatively show how the y-spins are converted to z-spins in the Fe:MoS$_2$ layer, we derive the expression for spin chemical potential distribution of z-polarized spins ($\mu_z(z)$). Based on the macroscopic model,[75] The spin-diffusion equation gives the partial differential equation relation of $\mu_k$ in Pt and Fe:MoS$_2$. Given a current flowing in the x direction of the Pt Hall bar, the relation of $\mu_y$ can be written as follows, based on the spin-diffusion equation:[75]

$$\frac{\partial_z^2 \mu_y}{\partial z^2} = \frac{\mu_y}{l_{sp}^2} \quad (8)$$

where $l_{sp}$ is the spin diffusion length in the Pt layer, and $\mu_y$ is the chemical potential of y-spins. For the chemical potential of spins in the Fe:MoS$_2$ layer, the contribution of spin-to-spin conversion due to SOC should be considered by adding terms including the crystal momentum of electrons (*k*):

$$\frac{\partial_z^2 \mu_y}{\partial z^2} - k_{y,z}^z \frac{\partial_z \mu_z}{\partial z} = \frac{\mu_y}{l_{sf}^2} \quad (9)$$

$$\frac{\partial_z^2 \mu_z}{\partial z^2} - k_{z,z}^y \frac{\partial_z \mu_y}{\partial z} = \frac{\mu_z}{l_{sf}^2} \quad (10)$$

Here $k_{i,j}^a$ is the crystal momentum of electrons with a-spin, where *i* denotes the direction of generated spin density and *j* represents the flow direction of spin current. and $l_{sf}$ is the spin diffusion length of Fe:MoS$_2$.

To obtain an approximate solution for Equations 8 and 9, we set up a model as shown in the inset of **Figure R1**. For a thin film of Fe:MoS$_2$ on the top of Pt, we assign z=0 at the Fe:MoS$_2$/Pt interface and solve the equation for 0<z<$l_{sf}$. We ignore the interfacial spin scattering by assuming the interface is transparent to spin tranmission. The boundary conditions can be expressed as: $\mu_z(z=0) = 0, \mu_y(z=0) = 1$ (normalized maximum $\mu_y$); $\mu_y(z=0+) = \mu_y(z=0-)$, $\mu_z(z=0+) = \mu_z(z=0-)$; $J_{Ps,z}^z(z=0) = 1$, and $J_{Fs,z}^z(z=0) = 0$. Given that the electrical conductivity of Pt is more than three orders of magnitudes larger than that of Fe:MoS$_2$ ($\sigma_P \gg \sigma_F$), the solution of $\mu_z(z)$ for 0<z< $l_{sf}$ can be expressed as:

$$\mu_z(z) \approx 2eE_x[l_{sf}^2 k_{y,z}^z \theta_p \left(\cosh\left(\frac{z}{l_{sf}}\right) - 1\right) + l_{sf} \theta_F \sinh\left(\frac{z}{l_{sf}}\right)] \quad (11)$$

where $e$, $E_x$, $\theta_p$ and $\theta_F$ represent elementary charge, the electronic field along the x-direction, spin Hall angle of Pt and Fe:MoS$_2$, respectively. The detailed distribution of z-spin density can be seen by plotting the $\mu_z(z)/k$ relation. The spin Hall angle of Fe:MoS$_2$ ($\theta_F$) can be estimated by critical current density ($J_c$) of SOT switching for thin magnet, given by:[76]

$$J_c \approx \frac{2e}{\hbar} \mu_0 M_s t_{FM} \theta_F \quad (12)$$

where $\mu_0$, $M_s$ and $t_{FM}$ are the magnetic permeability in vacuum, saturated magnetization, and thickness of Fe:MoS$_2$. Based on our previous literature report,[37] the $M_s$ at room temperature is about 65 emu cm$^{-3}$, and we take $t_{FM}$=0.9 nm, $J_c = 20 \times 10^6 \text{A} \cdot \text{cm}^{-2}$ (at 300K), the $\theta_F$ is approximately 0.17. Considering the $\theta_p$ for Pt is 0.068,[77] we plot the $\mu_z(z)$ versus ($z/l_{sf}^{\square}$) relation as shown in the blue curve of **Figure S11**. For $k_{y,z}^z$ is a positive constant smaller 1 for Fe:MoS$_2$, $\mu_z(z)$ increases monotonically for 0<z<$l_{sf}^{\square}$, implying the importance role of the thickness of Fe:MoS$_2$ layer for the spin (y)-to-spin (z) conversion

1.3 spin conversion in monolayer Fe:MoS$_2$
As shown in **Figure S11**, the dimension of $l_{sf}^{\square}$ is essential to verify whether spin-to-spin conversion is accountable for the SOT switching, as $\mu_z(z)$ decreases parabolically with a reduction of z. Based on Equation 10 and the previous studies on WTe$_2$,[11] the spin-to-spin conversion would be not efficient when z<$l_{sf}^{\square}$. Whereas a precise spin-diffusion length measurement was not conducted in this study, we estimated $l_{sf}^{\square}$ using the spin diffusion length of pristine MoS$_2$ ($l_{sm}^{\square}$). It is noteworthy that $l_{sm}^{\square}$ strongly depends on the number of layers of MoS$_2$, as a thinner layer will introduce more inversion symmetry breaking along the out-of-plane (z) direction. This symmetry breaking enables SOC to generate an equivalent k-dependent magnetic field via Dresselhaus interactions, leading to spin relaxation through the D'yakonov-Perel (DP) mechanism.[78,79] Consequently, for a 6-layer MoS$_2$, $l_{sm}^{\square}$ is ~200 nm below 30 K,[80] whereas $l_{sm}^{\square}$ is ~15 nm for a monolayer below 30 K.[81,82] At room temperature, the caculated $l_{sm}^{\square}$ for the monolayer is reduced to ~1.2 nm. [82]

Moreover, a stronger spin scattering has also been reported by Hanle-kerr and spin injection experiments on the monolayer MoS$_2$. The reported a spin lifetime ($\tau_{sm}^{\square}$) is shorter than 200 ps above 40 K [ref],[83,84] which can further reduce the spin diffusion length in monolayer at room temperature.
For Fe:MoS$_2$, Fe substitutional doping is expected to further decrease the spin diffusion length ($l_{sf}^{\square} < l_{sm}^{\square} \sim 1.2 \text{ nm}$) due to impurity scattering.[75,85] It is also important to note that the solution of $\mu_z(z)$ in Equation 9 is approximation that neglects spin scattering between interface of Fe:MoS$_2$/Pt. With the assistance of interface spin scattering, the actual value for $\mu_z(z)$ would be higher than our initial prediction in **Figure S11**.[11]

As a result of DP mechanism dominant spin scattering at high temperatures, enhanced interface spin scattering at 2D limit, and extra Fe doping-induced impurity scattering, <u>it is participated that the monolayer Fe:MoS$_2$ possesses a spin diffusion length smaller than 1.2 nm</u>. This suggests that the 0.9 nm thickness of monolayer Fe:MoS$_2$ (**Figure S1C**) is sufficient for the y to z spin conversion.

2. Modulation of R$_{THE}$ by pulsed current.
As shown in Figure S6 a-e, after applying a pulsed current ($J_p$), the $R_H$ signal was measured, showing a gradual decrease in $R_{THE}$ signal with increased current amplitude (Figure S6e). In contrast to AHE under d.c current, where both $R_{AHE}$ and $R_{THE}$ decrease proportionally with the temperature increase, $J_P$ changes $R_{THE}$ significantly while $R_{AHE}$ remains relatively stable.

3. Influence of DMI on magnetization switching
The influence of DMI on magnetization switching is quantified through the dynamics of magnetic layers, as described by the Landau-Lifshitz-Gilbert (LLG) equation:

$$\frac{\partial \boldsymbol{m}}{\partial t} = \frac{\gamma_0}{1+\alpha^2}\left(\boldsymbol{m} \times \boldsymbol{H}_{eff} + \alpha \boldsymbol{m} \times \boldsymbol{m} \times \boldsymbol{H}_{eff}\right) + \boldsymbol{\tau}^{SOT} \qquad (4)$$

where $\boldsymbol{m}$ represents the normalized magnetization vector, $\alpha$ denotes the Gilbert damping coefficient, $\gamma_0$ is the gyromagnetic ratio and $\boldsymbol{\tau}^{SOT}$ is the spin-orbit torque, which includes the torque arising from broken crystal symmetry, $\boldsymbol{\tau}_{AD}^{OOP}$. Since the effective magnetic field $\boldsymbol{H}_{eff}$ is proportional to the combination of magnetic stray field energy, perpendicular anisotropy energy, interfacial DMI energy and Heisenberg exchange energy [25,86], the $\boldsymbol{H}_{eff}$ can be expressed by $\boldsymbol{H}_{eff} \propto \boldsymbol{H}_{stray} + \boldsymbol{H}_{anis} + \boldsymbol{H}_{DMI} + \boldsymbol{H}_{exch}$. Among these components, the interfacial DMI field is directly proportional to the strength of interfacial DMI ($\boldsymbol{H}_{DMI} \propto D$). Therefore, the existence of interfacial DMI here can be a critical factor contributing to field-free SOT switching in the Fe:MoS$_2$/Pt heterostructure.

3. Origin of humps in Hall measurement.

**Figure S13A** presents Hall resistance (R$_H$) data measured at 4K, showing two sets of antisymmetric humps in the R$_H$ loop, one at ±2.5 kOe (hump a) and another at ±0.5 kOe (hump b), marked by blue and green arrows. As shown in **Figure S13A**, after exhibiting the hump at 0.5 kOe, the Hall signal does not return to the blue dashed lines, indicating there could be another ferromagnetic subdomain.

To explore the origin of the humps, we decomposed the Hall resistance ($R_H$) in **Figure S13A** using $R_H = R_0 H + R_{AHE} + R_{THE}$. We first subtracted the ordinary Hall effect ($R_0 H$) using a linear background, resulting in the red solid curve in **Figure S13.B**. A recurring observation in the literature is that the hump location typically coincides with the AHE domain reversal, which can be attributed to either THE[50,51,64–66] or two-channel AHE.[67,68] Accordingly, we further decomposed the Hall signal (red curve) into two channels with humps near the domain reversal. As shown in **Figure S13.B,** R$_{H1}$ (blue dashed line) presents a Hall loop with domain reversal near 'hump a' at ±2.5 kOe, while the remaining Hall signal, R$_{H2}$ (black dashed curve), shows a Hall loop with domain reversal near 'hump b' at ±0.5 kOe.

First we studied the R$_{H1}$ signal under two-channel AHE interpretation, by fitting the upper half of R$_{H1}$ with two $M_0 \tanh\left(\frac{H}{a_0} - H_0\right)$ functions.[50] As shown in **Figure S13.C**, R$_{H1}$ can be fitted by the superposition of two tanh functions, resulting in the solid red curve. However, because the peak location of R$_{H2}$ occurs before zero, this fitting results in one tanh function crossing the y-axis while x is positive (green solid curve), yielding a negative coercivity ( Hc =−1). This is clearly not allowed for the AHE response.

Next, we examined R$_{H1}$ signal under AHE + THE interperation, with one tanh function adding a hump from THE. In **Figure S13.D**, the solid red curve represents the fitted line for R$_{H2}$ data points, with the blue solid curve showing the AHE signal, where R$_{AHE}$ has an inverted sign and a coercivity of approximately 1 kOe. These fitting results suggest that it is unlikely that the hump *b* in **Figure S13.A** is attributed to the two-channel AHE. Furthermore, as shown in **Figure S9**, all hump features disappear when Pt is replaced with BST, suggesting that the Fe:MoS$_2$ crystal has homogenous magnetic order and doesn't fit with two-channel AHE model. Thus, we confirm that these "hump a" from R$_{H1}$ are likely to be related to THE.

For the $R_{H2}$ signal, the smaller 'hump b' can be explained by a THE signal at $\pm 2.5$ kOe appearing during the ferromagnetic domain reversals.[65] However, the 'hump b' might also be fitted with two groups of tanh function, with one function having y ($R_H$) sign inverted (see **Figure R3.C**) [87]. At this stage, the origin of 'hump a' suggests the presence of THE, but further spatial scanning of the magnetic vector is required to fully understand the origin of 'hump b'.